\documentclass[aps,prb, superscriptaddress,preprint,showpacs]{revtex4}

\usepackage{graphicx}
\usepackage{rotating}
\usepackage{longtable} 
\usepackage[usenames]{color}
\usepackage[normalem]{ulem}

\begin{document}

\title{
  {\em Ab initio} simulation of photoemission spectroscopy in solids: \\
  Plane-wave pseudopotential approach, with applications to
  normal-emission spectra of Cu(001) and Cu(111)
}

\author{Nata\v sa Stoji\' c}
\affiliation{SISSA -- Scuola Internazionale Superiore di Studi Avanzati, via
  Beirut 2-4, I-34014 Trieste, Italy}
\affiliation{  INFM-CNR Democritos, Theory @ Elettra group,
                       Trieste, Italy   }

\author{Andrea Dal Corso}
\affiliation{SISSA -- Scuola Internazionale Superiore di Studi Avanzati, via
  Beirut 2-4, I-34014 Trieste, Italy}
\affiliation{ INFM-CNR Democritos, Theory @ Elettra group,
                       Trieste, Italy   }

\author{Bo Zhou}
\affiliation{TASC National Laboratory, INFM-CNR, SS 14, km 163.5, I-34012
Trieste, Italy}

\author{Stefano Baroni}
\affiliation{SISSA -- Scuola Internazionale Superiore di Studi Avanzati, via
  Beirut 2-4, I-34014 Trieste, Italy}

\affiliation{ INFM-CNR Democritos, Theory @ Elettra group,
                       Trieste, Italy   }

\date{\today}

\begin{abstract}
  We introduce a new method for simulating photoemission spectra from
  bulk crystals in the ultra-violet energy range, within a three-step
  model. Our method explicitly accounts for transmission and
  matrix-element effects, as calculated from state-of-the-art
  plane-wave pseudopotential techniques within density-functional
  theory. Transmission effects, in particular, are included by
  extending to the present problem a technique
  previously 
  employed with success to deal with ballistic conductance in
  metal nanowires. The 
  spectra calculated for normal emission in Cu(001) and Cu(111) are in
  fair agreement with previous theoretical results and with
  experiments, including a newly determined spectrum. 
  The residual discrepancies between our results and the
  latter are mainly due to the well-known deficiencies of
  density-functional theory in accounting for correlation effects in
  quasi-particle spectra. A significant improvement is obtained by the
  LDA+$U$ method. Further improvements are obtained by including 
  surface-optics corrections, as described by Snell's law and
  Fresnel's equations.
\end{abstract}


\maketitle

\section{Introduction}

Photoemission spectroscopy (PES) is one of the most basic techniques
for investigating the electronic properties of
solids. \cite{Huf95,FeuWil76} In practice, however, it is difficult to
extract information directly from the observed spectra and 
theoretical considerations are necessary for a precise determination
of the underlying transitions.  The modeling of photoemission, as well
as the type and accuracy of the information that can be gained from 
experiments, depends on the energy range of the incident light. In the
energy range of ultra-violet photoemission spectroscopy (UPS, between 
$5$ and $100$~eV), the spectra are dominated by wavevector-conserving
transitions (direct transitions) with transition matrix elements differing
significantly for any pair of initial and final states.  Hence, in
UPS, final-state effects play a major role.  In empirical approaches
these final states are often modeled by free-electron bands, but, in
reality, they are influenced by the crystal potential especially at
low photon energies and, therefore, their proper description requires
detailed calculations.

Early photoemission calculations, ranging from
one-electron approaches to many-body
formulations,~\cite{Ada64,Mah70,SchAsh71,CarLedRou73,FeiEas74} covered various 
aspects of the photoemission process. One-electron photoemission
calculations started with the so-called three-step
model,\cite{BerSpi64} which breaks the photoemission process into
three independent steps: excitation of the photoelectron, its
transport through the crystal up to the surface, and its escape
into the 
vacuum. Inclusion of quasiparticle lifetimes through adjustable
parameters, within the multiple-scattering Green's function formalism,
led to the development of the one-step model.\cite{Pen76} Its modern
versions can model surfaces by a realistic
barrier\cite{GraBraBor93_II} and have the potential of replacing the 
previously adopted muffin-tin potentials by space-filling
potential cells of 
arbitrary shape,\cite{GraBraBor93} also taking into account
relativistic effects.\cite{GraBraBor94}

{\em Ab initio} methods based on density functional theory (DFT) are
nowadays considerably developed.\cite{Mar04,BarGirCor01} In particular, the plane-wave (PW)
pseudopotential (PP) formulation is being applied to a wide range of
properties and systems. Relativistic effects can be included in the PP
both at the scalar relativistic or at the fully relativistic level,
thus accounting for spin-orbit coupling.\cite{DalMos04} This
methodology, in principle, contains many of the ingredients necessary
to predict a photoemission spectrum from first principles, from which
information on various physical
properties of the system can then be extracted. In the case of X-ray
photoemission, for instance, the observed spectra are routinely
compared with the density of electronic states.  Considerable
complications, however, arise in the {\em ab initio} simulation of UPS
spectra, as well as in their interpretation, so that the application
of state of-the-art DFT PW-PP techniques to this problem has hardly
been attempted so far. First and foremost comes the difficulty of
accounting for the nonperiodic nature of the electronic states
involved in the photoemission process.  One of the few attempts to
compute photoemission spectra using PPs was made by Stampfl {\em et
al.}\cite{StaKamRil93} who constructed the final states by a
low-energy electron diffraction (LEED) computational technique.
Second, and no less important, is the well-known inability of DFT to
properly account for self-energy effects on the quasiparticle states
that are the main concern of PES. This failure of DFT to accurately
describe quasiparticle states is the field of intense research,
currently mainly addressed using techniques from many-body
perturbation theory, such as, {\em e.g.}, the GW approximation.
\cite{GWmethod,MarOniDel02,BotSchSol07}

In this paper the first problem is thoroughly addressed by calculating
the transmission of electrons from the crystalline medium into the
vacuum by a technique that was previously successfully employed to
deal with the ballistic conductance of an open quantum system within
the B{\"u}ttiker-Landauer approach.\cite{ChoIhm99} This technique,
originally formulated with norm-conserving PPs, has been generalized
to ultrasoft (US) PPs by Smogunov {\em et al.},\cite{SmoDalTos04} and
this generalization is used here to calculate the transmission into
the vacuum of the crystalline Bloch states. In addition to
transmission, a completely {\em ab initio} approach to PES would
require the calculation of dipole matrix elements and a proper account
of self-energy effects on the electron band structure, as well as of
the effects of the change of the dielectric function upon crossing the
surface (surface-optics effects). Dipole matrix elements are
calculated completely {\em ab initio} using a technique first
described by Baroni and Resta in 1986.\cite{BarRes86} The real part of
the self-energy shifts to DFT bands is accounted for semi-empirically
using the LDA+$U$ method, while the imaginary part (lifetime effects)
is simply added as an empirical parameter. Finally, surface-optics
effects are accounted for by the Snell's and Fresnel's
equations. While these equations could in principle be implemented
using a dielectric function calculated {\em ab initio}, for simplicity
we choose to implement them using experimental data.\cite{WeaKraLyn81}

As an application of our approach, we calculate the bulk contributions
to the normal photoemission spectra from Cu$(001)$ and Cu(111). Copper
is a prototypical system for UPS studies, for which many theoretical
results, as well as accurate experimental measurements, are available.
We compare our calculations with previous theoretical studies,
performed within the one-step and three-step models, and with
experimental data.  Not surprisingly, the main limiting factor in our
calculations appears to be the poor description of the Cu electron
bands by the local-density approximation (LDA), while transmission
effects are correctly accounted for, thus providing a viable way to
select and to weigh among the many available final conduction states
only those that couple to vacuum states.

The paper is organized as follows: in Sec.~II we describe the 
theoretical method used to calculate  photoemission spectra, 
while in Sec.~III we give some numerical details. In Sec.~IV we first
discuss the contributions of different terms in our expression for the
photoemission intensity, on one specific example; we then  present 
our {\em ab initio} results for the Cu(001) surface, 
obtained at the DFT level without empirical adjustments, 
followed by the results for Cu(001) and Cu(111) obtained 
from LDA+$U$ bands and accounting for surface-optics effects.
Sec.~V contains our conclusions.

\section{Theory}
In a three-step model, the photoemission current is proportional to the
product of the probability that an electron is excited from an
initial bulk state, $\psi_i$, to an intermediate bulk state,
$\psi_n$, of energies $E_i$ and $E_n$ and 
wavevector $\mathbf{k}$, $|M_{ni}(\mathbf{k})|^2$ (in this
transition the electron momentum is supposed to be conserved, in
spite of surface effects that break translational symmetry), times the
probability that the electron in the intermediate state is transmitted
into the vacuum, $T(E_n,\mathbf{k})$, conserving the
energy and the component of the momentum parallel to the surface,
$\mathbf{k}_\parallel$. Summing the composite 
probabilities all over the possible initial and intermediate states,
we obtain the current, $I$, as a function of the photo-electron
kinetic energy, $E_{\rm kin}$, and photon energy, $\hbar\omega$, using
the standard expression:\cite{GroEasFre79,SmiBenHur80,Huf95}
\begin{equation}
  I(E_{\rm kin},\hbar\omega,{\bf k_{\parallel}})\propto \sum_{ni}
  \int d k_\perp | M_{ni}({\bf k})|^2 T(E_n, {\bf
    k})\delta[E_n({\bf k})-E_i({\bf k})-\hbar\omega] 
  \delta[ E_{\rm kin} - E_n({\bf k}) + \Phi], 
  \label{eq:photoemission}
\end{equation}
where $k_{\perp}$ is the component of $\mathbf{k}$ perpendicular
to the surface and $\Phi$ the work function. The three-step model that we use is,
of course, an approximation which, in particular, does not properly
account for coherence between the excitation process occuring in the 
bulk and the escape of the electron, occuring at the surface. This coherence
may give rise to interference effects which are, therefore, neglected
in our approach. 
 
The transition matrix element in Eq. (\ref{eq:photoemission}), 
$M_{ni}$, is calculated from the interaction operator:
\begin{equation}
  H_{\rm int} = \frac{e}{2mc} ({\bf A \cdot p}+{\bf p \cdot A}),
  \label{eq:Hint}
\end{equation}
where  {\bf A} is the vector potential of the electromagnetic field and
{\bf p} the momentum operator of the electron, $e$ and $m$ are the
electron charge and mass, $c$ is the speed of light, and nonlinear
effects have been neglected. We will consider only the ${\bf A\cdot
p}$-type interaction, while the $\nabla A$-interaction, originating
from the second term in Eq.~\ref{eq:Hint} and giving rise to
surface emission (the gradient of {\bf A} is significant only
in a very narrow region around the surface), is neglected, in accord with common
practice in photoemission calculations.\cite{Pen76}
In this paper, we shall consider normal
photoemission only, {\em i.e.}  ${\bf k_{\parallel}}=0$. Energy conservation 
is imposed by the two $\delta$-functions.
In the dipole
approximation, {\bf A} can be considered spatially constant (the
wavelength of the photon beam, which is 120-2500~\AA~ in the UPS
energy range, is very large compared to the atomic spacing in crystal)
and therefore the transition matrix elements are proportional to the
dipole matrix element between the propagating initial and intermediate
states:
\begin{equation}
M_{ni} =\frac{e}{mc}{\bf A} \langle \psi_{n}|{\bf  p}|\psi_i\rangle.
\label{eq:Mfi}
\end{equation}
The vector potential {\bf A} carries information on the light 
polarization, depending on the polar ($\theta$, defined with respect 
to the surface normal) and azimuthal 
($\phi$) angles of incidence of the photon
beam. In this paper, we 
consider linearly polarized electromagnetic radiation, with the
following convention: for $p$-polarized light, {\bf A} is contained
within the plane formed by the directions of the incident light and
outgoing electron, while for $s$-polarized light, {\bf A} is
perpendicular to this plane. Thus, for $s$-polarized light
$\mathbf{A}$ is parallel to the surface for normal
photoemission. 

In general, when an
electromagnetic plane wave impinges on a metal surface, the value of
the vector potential transmitted inside the metal differs from the
value in the vacuum, due to the departure from unity of the medium
dielectric function, $\epsilon(\omega)$.\cite{WeaKraLyn81}
The transmitted vector potential ${\bf A}^t$  can be calculated from the
incident field ${\bf A}^i$, as described by Fresnel's equations,
derived by using the Maxwell's theory and Snell's law\cite{BorWol70}.
Actually, the difference between ${\bf A}^t$ and ${\bf A}^i$ can be
very large, especially for small photon energies ($\hbar\omega<20$~eV)
and large $\theta$, as extensively discussed in
literature.\cite{Fei74,Whi78,SmiBenHur80,GolRodFed83,WerCou84,WerCou85}

In order to calculate the transmission factor,
T($E_{n},{\bf k}$), we take into account that the
final state of the photoemission process is a time-reversed LEED
(TRL) state, which, sufficiently far from the surface is
free-electron-like in the vacuum ({\em outer region}), while inside
the crystal ({\em inner region}) it is a linear combination of the
Bloch states available at the intermediate energy, $E_n$. The TRL state can
thus be obtained by solving the one-electron Kohn-Sham (KS) Schr\"odinger
equation, subject to the appropriate boundary condition in the outer
region. This task is accomplished by matching the
wavefunctions in the inner and outer regions, using a method
proposed by Choi and 
Ihm,\cite{ChoIhm99} originally devised to cope with ballistic
conductance and later generalized to account
for US PP's.\cite{SmoDalTos04} In the outer region the TRL state is 
a plane wave whose wavevector has a
component perpendicular to the surface equal to
$k_\perp=\sqrt{2m E_{\rm kin}/\hbar^2 - {\bf k}_\parallel^2}$.
For given values of the photoelectron kinetic energy,
$E_\mathrm{kin}$, and parallel momentum, $\mathbf{k}_\parallel$,
in the inner region the TRL state reads:
\begin{equation}
\phi^{TRL}_{E_\mathrm{kin},{\bf k}_\parallel}({\bf r})=\sum_{m} 
\psi_n({\bf r}, {\bf k}_{m\perp}) t(E_n,{\bf k}_\parallel,{\bf k}_{m\perp} ),
\end{equation}
where the sum is over all the Bloch states
available at the intermediate energy
$E_n=E_\mathrm{kin}+\Phi$. In the intermediate region, 
the TRL depends on the details of the self-consistent potential at
the surface, and this dependence determines the relative amplitude
of the wavefunctions in the outer and inner regions, hence the
transmission coefficient.
In practice, the solution of the KS Schr\"odinger
equation by the method of Choi and Ihm\cite{ChoIhm99} provides
the coefficients of the expansion of the final photoemission state
in Bloch waves. 
These coefficients, which usually yield the total transmission and
hence the ballistic conductance, can be used to calculate the 
transmission probability into vacuum
$T(E_{n},{\bf k})= |t(E_{n},{\bf k})|^2$ separately
for each conduction band.  
We note that in this approach the scattering state is normalized in such a way
that both the incident plane wave and the Bloch states carry unit
current.

We now discuss the way in which  the two delta
functions appearing in Eq.~\ref{eq:photoemission} can be treated
in practice. The first delta function  imposes energy
conservation in the excitation step of the photoemission process, while 
the second one relates  the kinetic energy measured outside the crystal to 
the intermediate state energy, accounting for the work function. 
 The first delta function is usually represented as a Lorentzian:
\begin{equation}
\delta[E_{n}({\bf k})-E_i({\bf k})-\hbar\omega] = 
\frac{\Gamma_h/2 \pi}{[E_{n}({\bf k})-E_i({\bf k})-
\hbar\omega]^2 + \Gamma_h^2},
\label{eq:delta1}
\end{equation}
which corresponds to the spectral function of the 
hole left behind by the excitation process\cite{Mat98} and
results in the broadening of the initial state.  The width of the
distribution, $\Gamma_h$, gives the inverse lifetime of the hole, and
is equal to the absolute value of the imaginary part of the hole
self-energy. As in the majority of other photoemission studies,
we take $\Gamma_h$ as an adjustable parameter.  The second delta
function should be replaced by the analyzer resolution function, most
commonly expressed in the form of a Gaussian:
\begin{equation}
 \delta[ E_{\rm kin} - E_{n}({\bf k}) + \Phi] =
 \frac{1}{\sqrt{2\pi}\Gamma_{det}}\exp\Big[-\frac{1}{2}\frac{[ E_{\rm
     kin} - E_{n}({\bf k}) + \Phi]^2 }{\Gamma_{det}^2}\Big]. 
 \label{eq:delta2}
\end{equation}
$\Gamma_{det}$ is determined by the detector resolution, and is
related to the experimental energy broadening.\cite{Mat98}

We conclude by noticing that the bulk-emission model of photoemission
neglects electron damping caused by  the presence of a
surface. It is implicitly assumed in Eq.~\ref{eq:photoemission} that
${ k_{\perp}}-$conservation is perfect and the delta function for
this conservation law is omitted. The ${ k_{\perp}}-$conservation,
$\delta({k}_{n\perp} - {k}_{i\perp} - { G_\perp})$, is
usually represented by a Lorentzian, whose broadening parameter
describes damping.\cite{Mat98} The consideration of only bulk-emission
is a good approximation if the damping is small, {\it i.e.} if its
inverse, the electron escape length, $l_e$, is long enough, at least a
few lattice spacings. The escape length can be estimated from the relation:
$1/l_e = \Gamma_{n} \cdot \delta k_\perp/\delta E_{n} $,\cite{CouHufSch79}
where $\Gamma_{n}$ stands for the inverse lifetime of the intermediate state
and $\delta k_\perp/\delta E_{n}$ is the inverse group velocity of the
intermediate state. The escape length depends on the band structure through $\delta
k_\perp/\delta E_{n}$ term and on the photon energy through the
empirical dependence of $\Gamma_{n}$ on the intermediate-state
energy. Empirically one can use the relationship: $\Gamma_{n} = 0.065
\cdot E_{n}$.\cite{Mat98}

\section{Numerical details}

All the ingredients necessary to apply the theory
outlined in Sec.~II are 
calculated using DFT within the PW-PP approach, as implemented in
the {\tt PWscf} code of the Quantum ESPRESSO distribution.\cite{BarDalGir} 
In particular, the calculation of
transmission coefficients has been performed from the output of the {\tt
  PWcond} component contained therein.
For the exchange and correlation energy, we use the local density 
approximation (LDA) with the Perdew-Zunger parametrization.\cite{PerZun81} 
The interaction of the valence electrons with the nuclei and
core electrons is described by a Vanderbilt US PP.\cite{Van90} 
The use of the PP method for simulating PES deserves some
comments and requires care. Modern PPs are usually designed to
reproduce very faithfully the electronic structure (orbital energies
and one-electron wavefunctions outside the atomic core region) of
occupied states. Standard arguments based on the {\em transferability}
concept ensure that the quality of the electronic structure predicted
by PPs is as good in the energy range immediately above the Fermi
energy ($E_F$), {\em i.e.} in an energy range that extends up to, say, 10-15 eV
above $E_F$. In the present case,
however, a particular care has to be taken in describing the
intermediate state of the transitions, because these
lie at a higher energy 
than the transferability range of currently available PPs. For this
reason, we have decided to generate a highly accurate US PP, specially
designed for the purposes of the present work.
We used the $3d^{10}4s^14p^0$ atomic 
configuration of Cu, with the core radii: $r_s = 2.1$, $r_p = 2.4$, 
$r_d = 2.0$~a.u. and two projectors in each of the $s$, $p$, and
$d$-channels, one of which was chosen correspondingly to an atomic
state chosen at higher energy than usually done.\cite{PP_energies} 
The PP energy bands thus obtained agree within 0.05 to
  0.20~eV with those calculated from a highly accurate all-electron
  method, using the WIEN2k code,\cite{BlaSchMad01} up to 40 eV
above $E_F$. Ordinary Cu PPs,
generated without high-energy projectors, tend to miss some
unoccupied electron bands and have larger deviations from the
all-electron results with increasing energy.

To calculate T($E_{n},{\bf k})$, we used the self consistent
potential of a nine-layer tetragonal slab along the [001] direction
separated by a vacuum space equivalent to eleven layers. We chose a
vacuum space with length equal to two bulk layers as the unit cell in
the left lead and two central bulk-like layers of the slab as the
periodic unit cell of the right lead.  We used the experimental
lattice constant ($a=3.62~\mathrm{\AA}$) without relaxing the surface
layers.  However, our calculation allowing relaxations along the
perpendicular direction predicts $-2.8$~\% relaxation for the first
layer, $0.6$~\% for the second, and $0.2$~\% for the third.  We
checked that transmission factors change only negligibly with surface
relaxations in this case.  Kinetic energy cutoffs of $45$~Ry and
$450$~Ry have been used for the expansion of the wave functions and of
the charge density, respectively. These cutoffs, which are unusually
large for US PPs, are a consequence of the improved transferability
that we required from our custom-tailored PP. A Monkhorst-Pack
mesh\cite{MonPac76} $18\times 18\times 1$ was used for the slab
calculation. The band structure and the matrix elements of the dipole
operator were calculated from a bulk calculation in which
 $k_{\perp}$ was sampled by 860 {\bf k}-points.  The matrix elements of the dipole operator have
been compared to the matrix elements from the WIEN2K code, finding an
agreement of the order of 4\% for the states of interest in this
paper.  For the evaluation of the second delta-function from
Eq.~\ref{eq:delta2} we calculated the work function as the difference
between the Fermi level and the electrostatic potential in the vacuum,
and found $\Phi= 4.84$~eV, which is in agreement with previous
calculations\cite{FavDalBal01} and very close to the experimental
values ranging from $4.59$ to $4.83$~eV.\cite{work_fn_001} For
Cu(111), our LDA calculations gave the value of 5.08~eV, in good
agreement with a previous calculation (5.10~eV)\cite{PolMetSch93} and
experiment (4.9 and 4.94~eV)\cite{work_fn_111} The broadening
parameters which appear in Eq.~\ref{eq:delta1} and \ref{eq:delta2} are
chosen as $\Gamma_h=0.04$~eV and $\Gamma_{det}=0.07$~eV.

\section{Results}

In this section we first illustrate our method by discussing the
various contributions to the spectrum, as calculated from
  Eq. (\ref{eq:photoemission}), for Cu(001) at one specific
photon frequency ($\hbar \omega=23$ eV) and for one
specific angle of incidence ($\theta=65^o$) of the
incoming photon beam of $p$ polarization.  
We then present calculations for a more extensive
set of frequencies and  incidence angles. Finally, we
try to correct the two main sources of errors in our calculations by
studying how the spectra are modified by the LDA+$U$ bands and by
surface optics. The corrected spectra are presented also in the case
of Cu(111) surface.

\subsection { Illustration of the method}

\begin{figure}[h]
  \begin{center}
    \includegraphics[width=9.5cm]{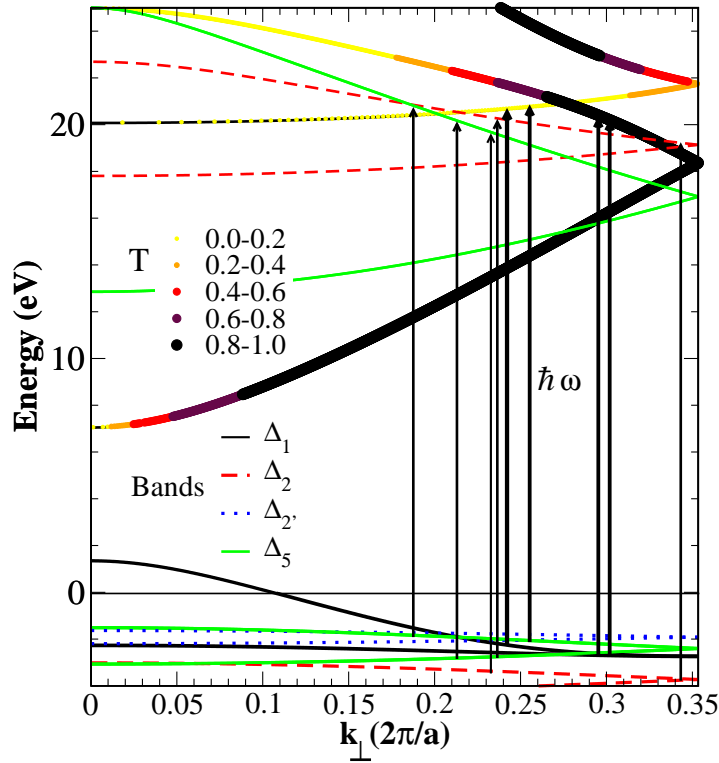}
    \caption{(Color online) Band structure of bulk Cu along the $k_{\perp}$ direction for the
      Cu(001) surface. We indicated, among the unoccupied bands, those that have
      a transmission probability into vacuum larger than 0. 
      Different point size and colors indicate different 
      transmission probability. The zero of the energy is at the Fermi level. 
      The direct transitions for $\hbar\omega=23$~eV are presented by arrows,
      while the four direct transitions which are simultaneously
      allowed and have non-zero transmission and dipole matrix element
      are shown with thick arrows.}
    \bigskip
    \label{fig:Trans001}
  \end{center}
\end{figure}

In Fig.~\ref{fig:Trans001} we show our calculated  energy bands of bulk Cu.  
The bands are plotted in the $\Gamma-Z$ direction, along [001]. These correspond 
to the bands along the $\Gamma-X$ direction in the fcc Brillouin zone (BZ) 
refolded in the tetragonal BZ. On top of the empty bands in  Fig.~\ref{fig:Trans001},
we add information regarding $T(E_n, \mathbf{k})$, wherever 
it is greater than 0. This is possible because $T(E_n, \mathbf{k})$ explicitly
depends on  $k_\perp$ on the intermediate states.\cite{note_T}
Thus, reading from  Fig.~\ref{fig:Trans001} at each
intermediate energy, we can find if the propagating final states exist ($T > 0$), 
and if so, for which values of the intermediate $k_\perp$.
In agreement with empirical intermediate-state 
determinations,~\cite{CouHuf84} we find that a free-electron-like
band properly modified by the ionic PP has the strongest
coupling to the vacuum state.  All the bands with nonvanishing
transmission probability belong to the $\Delta_1$ representation. This
is a result of the selection rules for normal photoemission, which
impose that the intermediate state be totally symmetric with respect to all
the $C_{4v}$ symmetry operations.  Combining this
result with the symmetry properties of the dipole matrix element we
obtain the allowed transitions for normal
photoemission from the (001) surface of an fcc crystal: for the
$z$-polarization only $\Delta_1 \rightarrow \Delta_1$ transition is
allowed, while for the $x$/$y$-polarization $\Delta_5
\rightarrow \Delta_1$ transitions are allowed.~\cite{Her77,Huf95}
Note, however, that some bands with $\Delta_1$ symmetry might not be
transmitted into vacuum, so symmetry alone would not be sufficient to
identify the intermediate states. In the same figure, we also display nine
direct transitions present in the case $\hbar \omega=23$~eV. 
Out of these transitions, only four satisfy selection rules and have
dipole matrix elements and transmission factors which are both 
non-zero. Two of these transitions are 
of the $\Delta_1 \rightarrow \Delta_1$ type and remaining two 
of the  $\Delta_5 \rightarrow \Delta_1$ type.  
Only one transition ($\Delta_1 \rightarrow \Delta_1$) has large 
transmission factor and large dipole matrix element,
while the other of the same type has small transmission factor.
Both transitions of the  $\Delta_5 \rightarrow \Delta_1$ type
have small dipole matrix element and one of them even a small transmission 
factor (smaller than 0.2).

\begin{figure}[h]
  \begin{center}
    \includegraphics[width=8.5cm, angle=0]{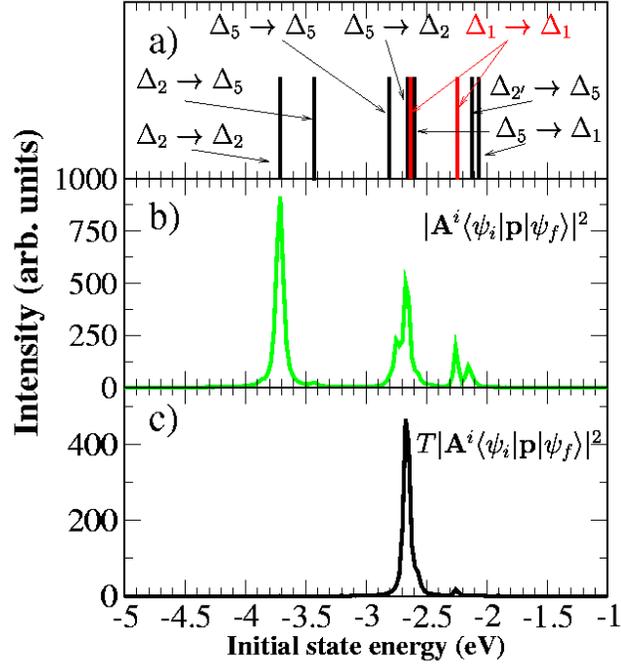}
    \hfill
    \caption{(Color online) Influence of dipole matrix elements and transmission
      factors on the spectrum for $\hbar\omega=23$~eV and
      $\theta=65^\circ$. Panel a)  presents all the allowed
      transitions with an indication of the symmetry of the initial
      and intermediate states. Panel b) shows the 
      effect of the dipole matrix
      elements. Panel c) shows the effects of the transmission factors.
    }
    \bigskip
    \label{fig:trans1}
  \end{center}
\end{figure}

In Fig.~\ref{fig:trans1} we illustrate the influence of the
dipole matrix elements and of the transmission factors on the shape 
of the spectrum. In panel {\em a} we show all the direct
transitions, regardless of 
the symmetry, their dipole matrix elements, and the transmission coefficients. 
In panel {\em b} we calculate the
photoemission spectrum setting the transmission factor to one for all the
bands. We find that the peaks with non-zero dipole matrix elements correspond
to $\Delta_2 \rightarrow \Delta_2$ and  $\Delta_1 \rightarrow \Delta_1$ 
($\Delta_2 \rightarrow \Delta_5$,  $\Delta_5 \rightarrow \Delta_5$, 
$\Delta_{5} \rightarrow \Delta_2$, $\Delta_{2'} \rightarrow \Delta_5$ and 
$\Delta_5 \rightarrow \Delta_1$) transitions 
for $z$-polarization ($x$/$y$-polarization), although 
the transitions with $\Delta_5$ and $\Delta_2$ intermediate states  
are not allowed for the normal emission. 
Including the dipole matrix elements, we obtain a spectrum in which
the intensity of the peaks may change. The polarization and the direction 
of the incident photon beam now play an important role. 
For $\theta=65^\circ$ the $z$-polarized transitions are enhanced
with respect to transitions due to $x$/$y$-polarized light. 
However, neglecting the probability of the intermediate 
states to be transmitted
to vacuum we still have many transitions into $\Delta_2$ and
$\Delta_5$ intermediate states which have finite 
intensity. Also, the relative
intensities of peaks with $\Delta_1$ intermediate states are incorrect. 
The introduction of the transmission factor not only selects the 
intermediate states with $\Delta_1$ symmetry, but also modulates the peak intensities.
Thus, two peaks shown in panel c) originate from the $\Delta_1$
initial state, while the shoulder of the main peak on the high-energy side and almost
invisible shoulder to the high-energy peak originate from the $\Delta_5$ initial state.
We note that, at variance with the rest of the paper, in Fig.~\ref{fig:trans1} 
we used smaller broadening parameters $\Gamma_h=\Gamma_{det}=0.015$~eV, 
in order to separate the different peaks. 

\begin{figure}[h]
  \begin{center}
    \includegraphics[width=8cm, angle=0]{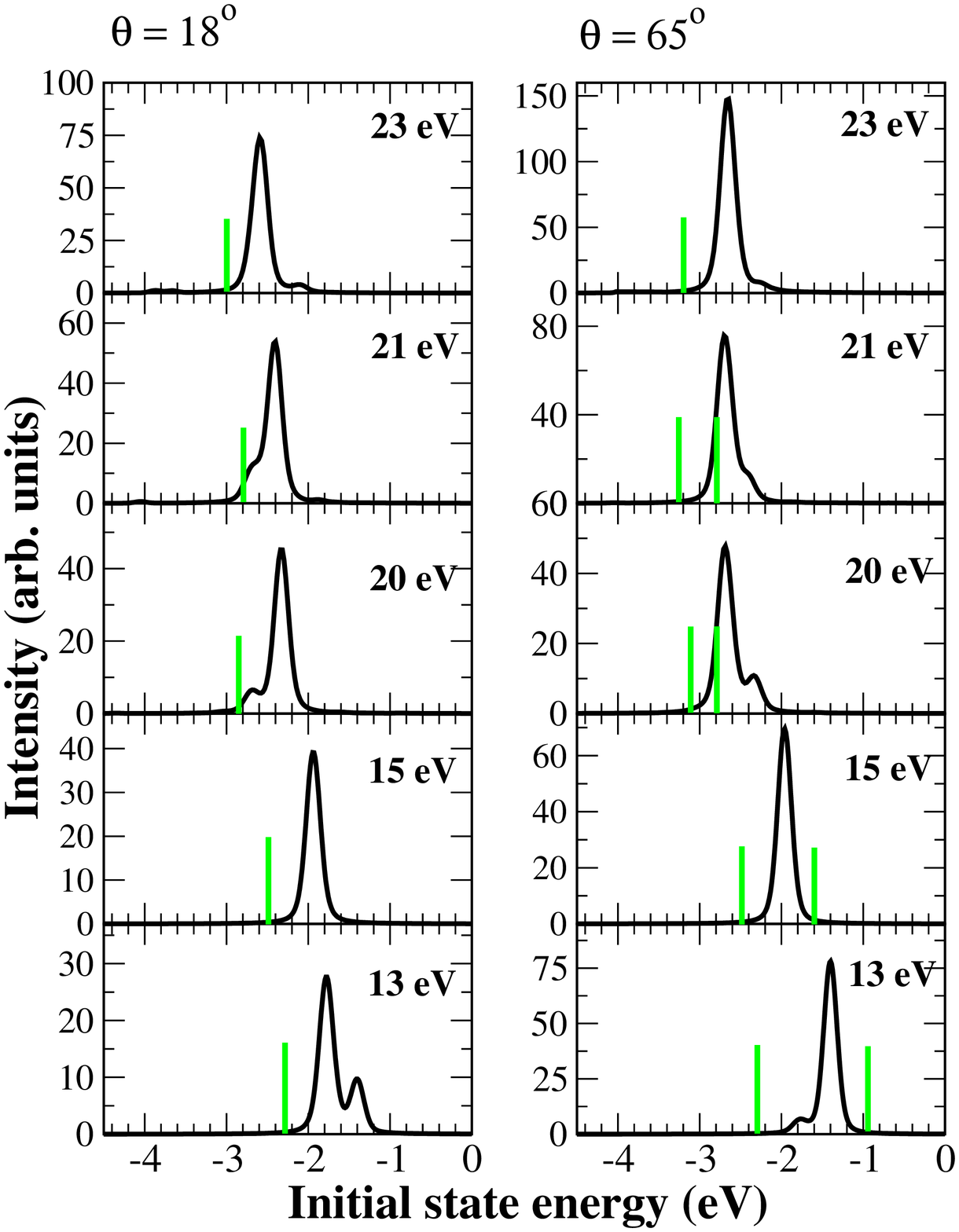}
    \hfill
    \includegraphics[width=8cm, angle=0]{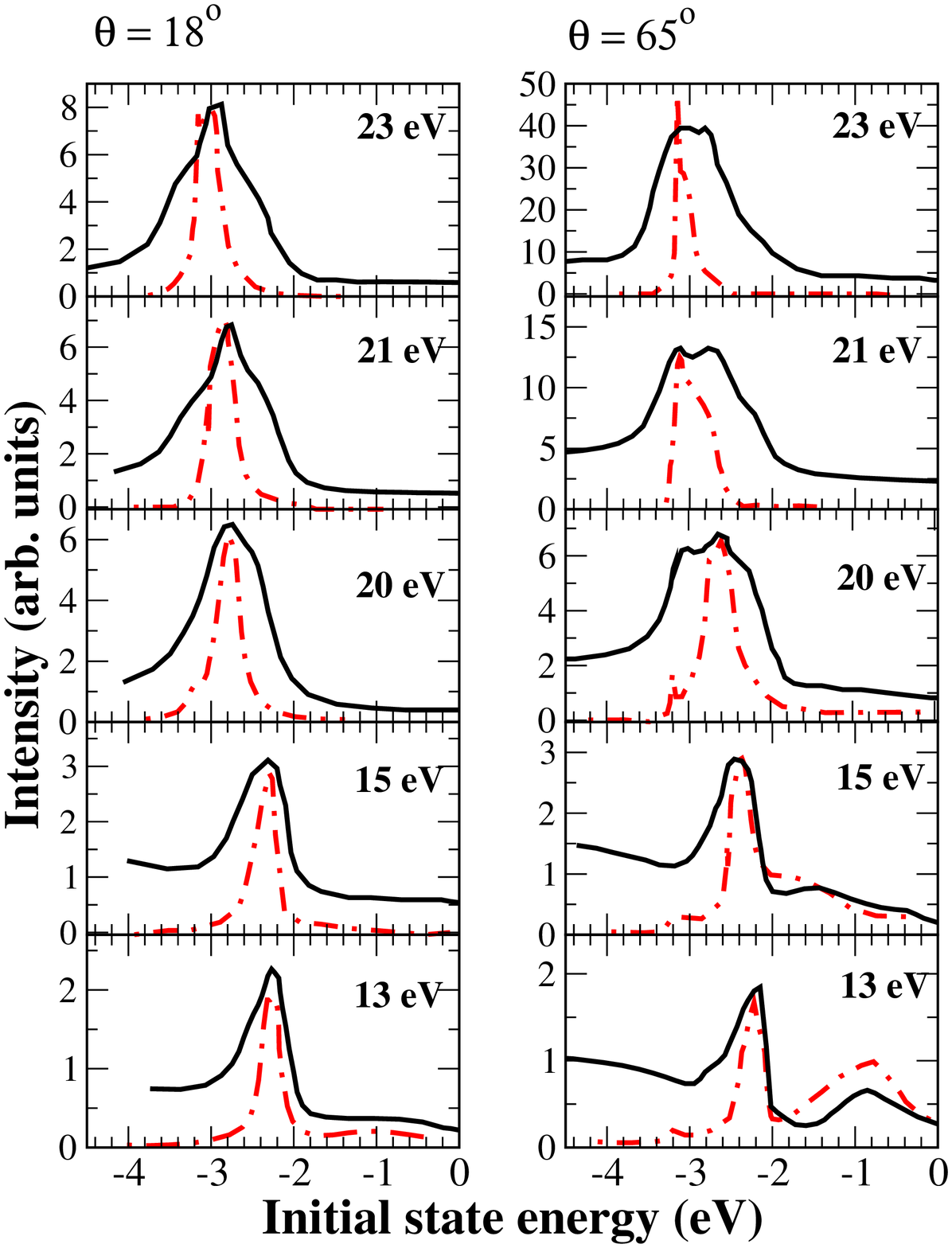}
    \caption{ (Color online)
      Calculated photoemission intensity for Cu(001) (left
      panels). The experimental spectra (dashed line), together with a
      previous calculation (darker full line)\cite{BaaBraDie86} are
      shown on the right. The spectra in both panels are given for
      various photon frequencies and  for two incident angles,
      $\theta=18^{\circ}$ (on the left of each panel) and
      $\theta=65^{\circ}$ (on the right). The sticks in the
      left panels indicate positions of the main experimental peaks.
    }
  \bigskip
  \label{fig:spectra001}
  \end{center}
\end{figure}

\subsection{{\em Ab initio} results}

In the left panel of Fig.~\ref{fig:spectra001} we present our
calculated spectra for five photon frequencies between $13$ and
$23$~eV and two incident angles $\theta=18^{\circ}$ and
$\theta=65^{\circ}$ for the $p$-polarization. In the right panel,
we show for comparison the experimental and previous theoretical
normal-emission spectra for the same frequencies and
angles.\cite{BaaBraDie86} Previous calculations were performed within
the one-step model,\cite{Pen76} based on the non-relativistic
empirical muffin-tin potential and taking into account the surface
optics through the application of Snell's law and Fresnel's
equations. Our spectra in Fig.~\ref{fig:spectra001} do not include 
surface-optics effects and are calculated within the bulk photoemission
model. The first approximation is quite severe and its effects will be
discussed below, while the second should be sufficiently justified for
this surface. Actually, for the photon energies considered here,
$\delta k_\perp/\delta E_{n}$ is smaller than $0.05$~$\mathrm{\AA}^{-1}
\mathrm{eV}^{-1}$ and the damping (for an average photon energy of
$18$~eV) can be estimated to be less than 0.06~$\mathrm{\AA}^{-1}$.
This rough estimate yields the electron escape length of about $10$
lattice spacings in the direction perpendicular to the surface.  The
choice of the incidence angle influences the spectra significantly:
for $\theta=18^{\circ}$, light is mostly polarized in the $xy$
plane, (the initial state belongs to the $\Delta_5$ representation),
while for $\theta=65^{\circ}$, most emission is from
$\Delta_1$-like states ($z$-polarization).  Overall, our calculation
reproduces the majority of the experimental peaks, albeit with a shift
of about $0.2$-$0.6$~eV towards higher energies and somewhat altered
relative intensities. It is well known that the LDA, as well as generalized
gradient approximation (GGA), fails
to describe accurately quasi-particle energy bands 
as measured by PES and, therefore, the incorrect position of
the peaks has to be attributed to the error in the calculated
bands.\cite{StrClaNic98} The calculation within the one-step 
model\cite{BaaBraDie86} is performed with an empirical potential and, 
therefore, the energy bands correspond to the experimental 
bands rather well.  The error in the
positions of energy bands can result also in a reduction or an
increase of the number of peaks.  At $\theta=18^{\circ}$ and
$\hbar \omega = 13$~eV the theoretical spectrum shows two peaks, one
at $-1.76$~eV and one at $-1.38$~eV. The former is due to a transition
from the $\Delta_5$ band, while the latter originates from the
$\Delta_1$ band.  Experimentally, only a single peak is present at
about $-2.30$~eV.  The spectrum for $\hbar \omega = 15$~eV shows a
single peak as the experiment although at higher energy.  The spectra
at $\hbar \omega = 20,21$ and $23$~eV show only one main peak and a
small peak, missing some of the shoulders present in the experimental
spectra, a feature that our result has in common with the previous
calculation. The lower-energy features in the spectra $\hbar \omega = 20$ and 21~eV
originate from the $\Delta_1$ bands and the higher-energy features from the
$\Delta_5$ bands.  In the spectrum $\hbar \omega = 23$, both peaks are predominantly 
from the $\Delta_5$ initial state, although there are significant contributions on the
low-energy sides originating from the  $\Delta_1$ initial states. These contributions
are hard to discern due to a
large broadening and closeness of the peaks ($\sim 0.1$~eV).

For $\theta=65^{\circ}$ the agreement is somewhat
worse. The spectrum for $\hbar \omega = 13$~eV
has a barely visible feature in place of the main peak of the
experimental spectrum while the $\Delta_1$ peak is
significantly overestimated and at too low energy.  As we show below,
this will be corrected in part by considering the surface optics. That
type of correction is quite large for small photon energies and large
incidence angles.\cite{SmiBenHur80}  Similarly to the case
$\theta=18^{\circ}$, the spectrum for $\hbar \omega = 15$~eV has
only one peak, which actually contains two transitions (from the
$\Delta_1$ and $\Delta_5$ initial states).  Due to our imprecise
energy bands, both transitions are accidentally at the same energy.
The experimental spectrum for $\hbar \omega = 20$~eV has two peaks of
roughly equal intensity, with a broad shoulder on the high-energy
side, while our spectrum reproduces just one main peak and a much
lower-intensity peak at higher energy. Our spectra for $\hbar \omega =
21$ and $23$~eV are underestimating or missing a high-energy feature,
present in the experimental spectra. We note that in contrast to the
$\theta = 18^\circ$ case, both peaks in the spectrum for $\hbar \omega = 23$~eV
originate from the $\Delta_1$ initial states, and transitions from the
$\Delta_5$ give small contributions  on the high-energy sides, as seen
in Fig.~\ref{fig:trans1}.
Finally, we note that the peak intensities can be
compared only within one spectrum, {\em i.e.} the intensities for
different $\hbar\omega$ cannot be compared in our calculation, due to
the neglect of electron damping, which is energy dependent.
Consequently, it is clear that the standard {\em ab initio} approach
needs further corrections to reproduce the fine details of the
experimental spectra.

\subsection{Empirical corrections}

In this section we try to correct empirically two of the main shortcomings 
of our approach using quite simple models. 
It is known that inclusion of self-energy effects, at least within
the GW model, would be mandatory for a realistic description of the
band structure.\cite{MarOniDel02} 
Presently, however, this is beyond our capabilities mainly because it 
would require a nontrivial extension of the ballistic conductance code.
Hence, we choose a simpler approach, using LDA+$U$.\cite{AniAryLic97,CocGir05}
LDA+$U$ goes beyond LDA by treating exchange and 
correlation differently for a chosen set of states, in this case, 
the copper $3d$ orbitals. The selected orbitals are treated with 
an orbital dependent potential with associated effective on-site Coulomb 
interaction $U_{\rm eff}$, which is a function of Coulomb and exchange interactions $U$ and $J$,
 $U_{\rm eff}=U - J$.~\cite{SawMorTer97,DudBotSav98}
The LDA$+U$ method is most commonly known as a cure for
  the inability of traditional DFT implementations to predict the
  insulating state of some strongly correlated materials.\cite{AniAryLic97}
  Although the theoretical
  foundation of LDA$+U$ is somewhat questionable, its range of
  applicability is wider, and this method has indeed been succesfully
  applied to metallic systems where the effects of electron correlations are
  intermediate.\cite{MohPerBla01,StoBinAlt06,StoBin08} LDA$+U$ is also being
  succesfully employed as a predictive tool in the chemistry of
  transition-metal molecules.\cite{KulCocSch06} Furthermore, in
  the specific case of bulk copper, there is evidence that an account
  of self-interaction effects in LDA through the LDA+SIC approach
  leads to  an improvement of the calculated bands.\cite{Nor84}
  However, the LDA+SIC approach neglects screening
  effects on the self-interactions, which are instead accounted for to
  different degrees of accuracy in the GW and in the LDA+$U$
  methods. While GW addresses screening in a more rigorous way, LDA+$U$ can be
  considered as the static limit of a kind of (admittedly, rather crude)
  approximation to the GW method.\cite{AniAryLic97}

\renewcommand{\baselinestretch}{1}
\begin{table}[ht]
\bigskip
\begin{center}
\begin{tabular}{ c c c c c c c c c}
\hline
\hline
  &  &  LDA &  GW & LDA+$U_1$ & LDA+$U_2$ & LDA+$U_3$ & Experiment \\
\hline
 Positions of & $\Gamma_{1,2}$              & $-2.18$ & $-2.81$ &$-2.60$ & $-2.75$ & $-2.92$ & $-2.78$ \\
 $d$ bands    & $X_5$                       & $-1.44$ & $-2.04$ &$-1.85$ & $-2.00$ & $-2.17$ & $-2.01$ \\
              & $L_3$                       & $-1.60$ & $-2.24$ &$-2.00$ & $-2.17$ & $-2.31$ & $-2.25$ \\
\smallskip
              & $\Gamma_{1,2}-\Gamma_{2,5}$ & $0.83$  & $0.60$  &$0.83$  & $0.83$  & $0.83$ & $0.81$ \\
 Widths of    & $X_5-X_3$                   & $2.94$  & $2.49$  &$3.01$  & $3.04$  & $3.05$ & $2.79$ \\
 $d$ bands    & $X_5-X_1$                   & $3.40$  & $2.90$  &$3.53$  & $3.57$  & $3.61$ & $3.17$ \\
              & $L_3-L_3$                   & $1.44$  & $1.26$  &$1.47$  & $1.48$  & $1.49$ & $1.37$ \\
              & $L_3-L_1$                   & $3.46$  & $2.83$  &$3.55$  & $3.56$  & $3.60$ & $2.91$ \\
\smallskip
 Positions of & $\Gamma_1$                  & $-9.37$ & $-9.24$ &$-9.25$ & $-9.22$  & $-9.19$ & $-8.60$ \\
$s/p$ bands   & $L_2'$                      & $-1.00$ & $-0.57$ & $-0.88$ & $-0.85$ & $-0.83$ & $-0.85$ \\
\smallskip
 $L$ gap      & $L_1-L_2'$                  & $4.04$  & $4.76$  & $4.67$   & $4.88$ & $5.10$ & $4.95$ \\ 
\hline
\hline
\end{tabular}
\caption{ 
  Comparison of different theoretical (LDA, LDA+GW
  corrections,\cite{MarOniDel02} LDA+$U$, $U_1=1.5$~eV, $U_2=2$~eV and
  $U_3=2.5$~eV)  band energies and bandwidths for copper, at
  high-symmetry points, compared to the average over several
  experiments (as reported in Tables I and XIII in
  Ref.~\onlinecite{CouHuf84}). All values are in eV.
} 
\label{tab:ldaU}
\end{center}
\end{table}

As Cu has an almost completely filled $d$-shell, the main effect of the 
LDA+$U$ is the shift of the electron bands, while the eigenfunctions
are expected to remain quite close to the LDA ones.\cite{AniAryLic97}
Actually, we checked that, for the values of $U$ used here, the
overlap between the LDA and the LDA+$U$ wavefunctions is 
of the order of 0.99. Thus, we kept the same
transmission and dipole matrix elements calculated with the LDA
wavefunctions correcting only the band structure.

Table~\ref{tab:ldaU} presents some results, such 
as the positions of $d$-bands and bandwidths, evaluated using different 
methods including DFT with LDA, self-energy corrections within GW 
approximation\cite{MarOniDel02} calculated on top of {\em ab initio} 
DFT results, LDA+$U$ for three values of $U_{\rm eff}$, all compared 
to the average over several experimental values.\cite{CouHuf84}
The positions of the $d$-bands at the $\Gamma$ point vary greatly for 
different methods. The LDA calculation finds the band $0.6$~eV too shallow, 
while GW reproduces the experimental value quite well. LDA+$U$ significantly 
improves with respect to the LDA value and for $U_{\rm eff}=2$~eV gives 
almost the experimental value.
Similar level of precision can be seen for the positions of the $d$-bands 
at the $L$ and $X$ points, with  somewhat larger deviations of the LDA and 
the LDA+$U$ from the experimental value at the $L$ point. Again, among the three values of 
$U_{\rm eff}$, at the $L$ and $X$ points the best agreement 
with experiment is obtained for $U_{\rm eff}=2$~eV. 
The width of the $d$-band at the $\Gamma$ point is, instead, quite faithfully 
reproduced both by the LDA and the LDA+$U$. For other special points given in 
Table~\ref{tab:ldaU}, the widths remain almost constant for different $U_{\rm eff}$. We note 
that also the positions of $s/p$-bands and $L$-gap improve with the LDA+$U$. 
Overall, we conclude that the LDA+$U$ can correct the LDA bands in a significant manner, 
and has effects comparable to the full self-energy calculation. Also, on the 
basis of comparison with the experimental results, we find that 
$U_{\rm eff}=2$~eV gives the best results and we choose to use 
this value in the rest of the paper.

The second main problem  in the calculation of the intensities of the
photoemission peaks comes from the fact that the vector potential 
inside the solid is different from the vector potential in the vacuum.
Consequently, one should correct the intensities using the Snell's law
and Fresnel's equations.\cite{Fei74,Whi78,SmiBenHur80,GolRodFed83,WerCou84,WerCou85} 
An accurate account of this effect is quite difficult.
First of all, one should use a dielectric function
calculated consistently within the same {\em ab initio} scheme used
for the calculation of the other 
quantities. Furthermore, the effect of the surface should be properly taken
into account in the evaluation of the dielectric function. However, as this would 
require a substantial effort,  we choose just
to estimate the effect by using the experimental dielectric
function from Ref.~\onlinecite{WeaKraLyn81}.

\begin{figure}[h]
  \begin{center}
    \includegraphics[width=8cm, angle=0]{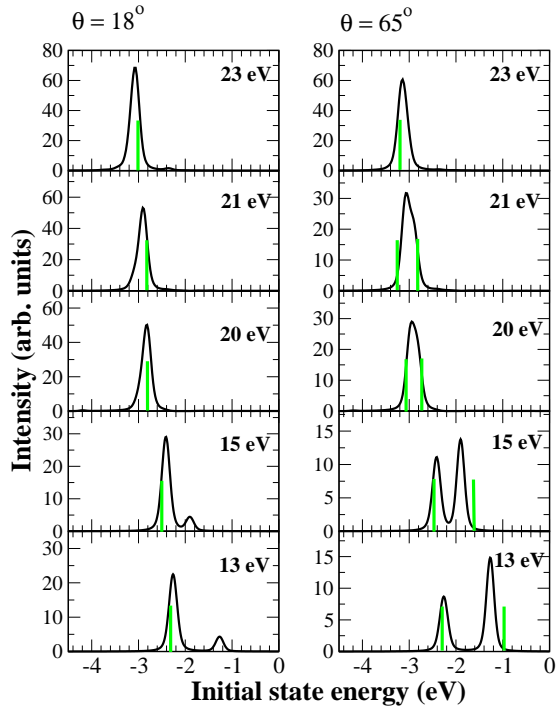}
    \hfill
    \caption{ (Color online)
      Calculated photoemission intensity for various photon
      frequencies for two incident angles, $\theta=18^{\circ}$ 
      and  $\theta=65^{\circ}$ using LDA+$U$ approach, 
      with $U_{\rm eff}=2$~eV.  The sticks indicate the
      positions of the main peaks in the experimental spectra (and
      previous calculations spectra).
    } 
    \bigskip
    \label{fig:ldau_001}
  \end{center}
\end{figure}

We report in Figure~\ref{fig:ldau_001} the photoemission spectrum calculated 
with the LDA+$U$ bands with $U_{\rm eff}=2$~eV and including the surface 
optics corrections, for the same parameters as in Fig.~\ref{fig:spectra001}. 
We checked that the spectra do not show strong dependence on the choice of 
$U_{\rm eff}$, e.g. for $U_{\rm eff}=1.5$~eV, the spectra are almost 
identical to the $U_{\rm eff}=2$~eV case, with a small shift in energy. 

In comparison with the experimental spectrum from Fig.~\ref{fig:spectra001}, 
we see that the energy positions are mostly corrected. For 
$\theta=18^\circ$, the spectrum for $\hbar\omega=13$~eV is improved 
with respect to the LDA spectrum (Fig.~\ref{fig:spectra001}), also in terms 
of the distance and relative intensities of the two peaks. 
For $\hbar\omega=15$ eV, the two transitions which were at the same initial
energy split, giving rise to a new peak. Thus, the lower-energy peak originates
from the $\Delta_5$ band and the higher-energy one from the $\Delta_1$ band. 
In the one-step 
calculation there is only a hint of a shoulder on the high-energy side. 
For $\hbar\omega=20$ and $21$~eV we lose the shoulders originating from the 
$\Delta_1$ initial state, present in the LDA spectra. This is because 
the bands are shifted in such a way that the transitions happen for
${ k_{\perp}}$ at which 
the two bands are very close in energy, less than 0.2~eV difference. 
The spectrum for $\hbar\omega=23$~eV looses a higher-energy peak originating from
the $\Delta_1$ band, as this peak comes under the main peak from the $\Delta_5$ band.

For $\theta=65^\circ$ all the spectra have significantly lower
intensity with respect to the LDA, as a direct consequence of
surface-optics corrections.  In the spectrum for $\hbar\omega = 13$~eV, the positions 
of the peaks are improved and also the peak intensity ratio goes in the
right direction, as a result of the inclusion of $\epsilon(\omega)$.
Nevertheless, this correction does not suffice and the intensities of the
two peaks remain wrong, both with respect to experiment and with respect
to the one-step model. Inspecting the Cu$(001)$ band structure in 
Fig.~\ref{fig:Trans001} we see 
that for $\hbar\omega=13$~eV the two initial bands actually have different 
dispersions at the ${ k_{\perp}}$ at which transitions are taking place (0.17 and 0.18~$2\pi/a$), 
whereas the dispersion of the intermediate band 
(the lowest unoccupied band) does not change significantly 
between the two ${ k_{\perp}}$. 
The low-energy peak, which is underestimated, originates from the initial 
band of $\Delta_5$ symmetry and has smaller slope at the ${ k_{\perp}}$ of 
the direct transition. Allowing nondirect transitions, due to electron 
damping, this peak would get many more contributions than the peak 
originating from a $\Delta_1$ band with strong dispersion and would, thus, 
improve agreement with the experimental spectrum. 
A similar argument holds also for the spectrum for 
$\hbar\omega=15$~eV which  gets a new peak with respect to the 
LDA spectrum, but whose intensity is overestimated. 
For $\hbar\omega=20$ and 21~eV the high-energy peaks (of $\Delta_5$ origin) from the 
LDA spectra are smeared 
into shoulders, due to the shift of bands. For the same reason,  a high-energy
shoulder from the LDA spectrum for  $\hbar\omega=23$~eV is lost.
Overall, we conclude that the LDA+$U$ correction is affecting 
all the spectra, causing the shift of all peaks, which also results in a decrease or increase
of the number of peaks. The surface-optics corrections improve agreement for low photon energies
and have stronger effects for $\theta=65^\circ$.  In general, both corrections
improve the agreement with experiment.

\begin{figure}[h]
  \begin{center}
    \includegraphics[width=8cm, angle=270]{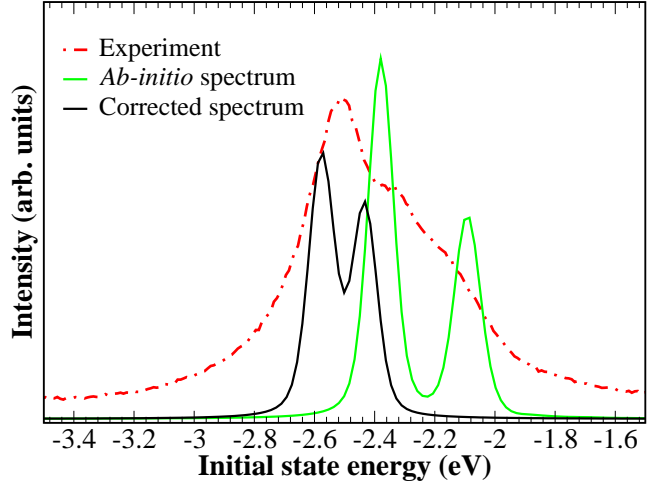}
    \hfill
    \caption{ (Color online)
      Comparison of our Cu(001) {\em ab initio} (light-colored full line) and
      corrected (dark full line) spectra, with the
      experimental spectrum (dashed line) for $\hbar
      \omega=17$~eV and $\theta=45^\circ$. 
    }
    \bigskip
    \label{fig:elettra}
  \end{center}
\end{figure}

Figure~\ref{fig:elettra} compares our {\em ab initio} and corrected 
spectra, with a recently measured experimental spectrum 
on the Cu(001) surface for $p-$polarization, $\hbar \omega=17$~eV and $\theta=45^\circ$.
The experimental spectrum was measured on the Cu(001) single crystal surface at APE beamline
(TASC, Italy)  at room temperature. It was integrated
over an angular window of $1^\circ$ around the normal emission. 
The energy resolution was estimated to be 25~meV.
Both theoretical spectra have two peaks, the low-energy one originates from
the $\Delta_1$ initial band and the high-energy one from the $\Delta_5$ initial
band. 
The {\em ab initio} spectrum has wrong energy positions ($\sim 0.15$~eV 
too high), the distance between the two peaks is too large and the peak 
ratio is overestimated ($1.78$ instead of $1.40$). 
The corrected spectrum shows a better agreement with experiment. The 
positions of the peaks are closer to the experiment ($\sim 0.07$~eV too deep),  
the distance between peaks is correct while the peak ratio is somewhat 
underestimated ($1.23$). However, we cannot reproduce the high-energy broad
structure present in the experimental spectrum. 
The experimental energy resolution  and inverse lifetime of  
the electron hole (estimated to be $\Gamma_h = 0.006\cdot E_i^2 + 0.01$~eV 
$\approx 0.04$~eV)\cite{Mat98} cannot account for the discrepancy between 
theory and experiment. Also, for this photon energy, broadening due to finite 
electron escape length should not be pronounced. 
We see that not all the peaks can be explained by direct transitions only 
and assuming a $\Delta_1$ intermediate state, as imposed 
by the selection rules for normal photoemission. We
note  that in Eq.~\ref{eq:photoemission} we disregarded the 
delta function describing the ${\bf k_{\parallel}}$ conservation. 
Performing analysis similar to the one in Fig.~\ref{fig:trans1}, we have 
found that there are two direct transitions, forbidden by selection rules, 
$\Delta_{5}\rightarrow \Delta_5$ and
$\Delta_{2'}\rightarrow \Delta_5$, located at $-2.41$ and $-2.38$~eV
in the corrected spectrum, respectively. They, also taking into 
account our somewhat imprecise energy positioning,
might correspond to the missing peak. They have a 
very large dipole matrix elements and it seems likely that
even a small in-mixing of these transitions might result in an observable structure 
in the photoemission intensity. The 
finite acceptance angle of the electron detector means that electrons are 
collected from a finite part of the surface Brillouin zone (broadening of 
${\bf k_{\parallel}}$). 
This implies that in the normal emission spectrum it is possible to 
have small contributions from the dipole-selection forbidden transitions.
These issues, however, are left for future investigations.

\begin{figure}[h]
  \begin{center}
    \includegraphics[width=8cm, angle=0]{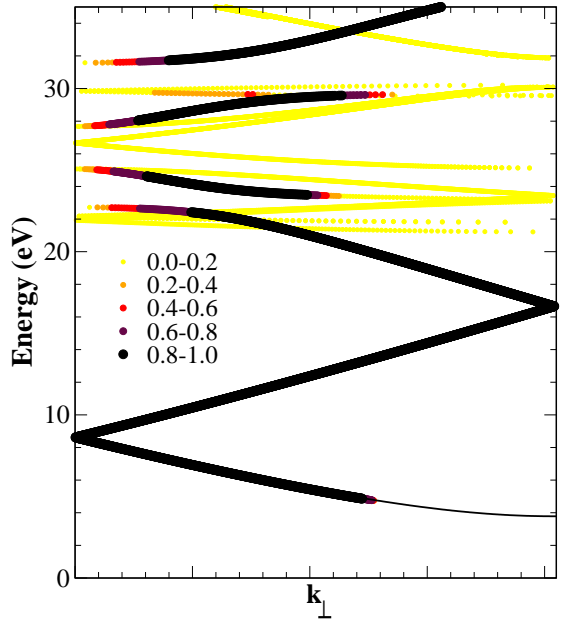}
    \caption{  
      Transmission intensity plotted on top of the unoccupied electron
      bands along ${ k_{\perp}}$ direction for Cu(111). Symbol size
      and color coding for transmission intensity is given in the
      legend. The Fermi level is at zero energy.  
    } 
    \bigskip
    \label{fig:Trans111}
  \end{center}
\end{figure}

{\bf{Cu(111)}}
As in the case of the (001) surface,  also for the (111) surface we  present our 
{\em ab initio} calculation of the transmission factors, 
given on top of the empty initial bands, and plotted versus the wavevectors perpendicular 
to the surface (corresponding to the $\Gamma-L$ line 
in the fcc cell),  in Fig.~\ref{fig:Trans111}. For this surface orientation, the unit cell 
has three atoms and the symmetry corresponds to the  $C_{3v}$ point group. Selection 
rules allow only $\Lambda_1 \rightarrow \Lambda_1$ transitions for the case of $z$-polarization 
and $\Lambda_3 \rightarrow \Lambda_1$ transitions for the case of $x$/$y$-polarization.
From Fig.~\ref{fig:Trans111} we conclude that for photon frequencies below 20~eV, transmission 
will be close to 1 for all dipole-allowed intermediate states.
Using the same reasoning as for the Cu(001) case,
we find that for the average photon energy of 8~eV, $k_{\perp}$ broadening is 
about $0.04~\AA^{-1}$. Our rough estimate yields the electron escape length of about 15 
lattice spacings, which ensures that the bulk model can be applied also in this case.

\begin{figure}[h]
  \begin{center}
    \includegraphics[width=15cm, angle=0]{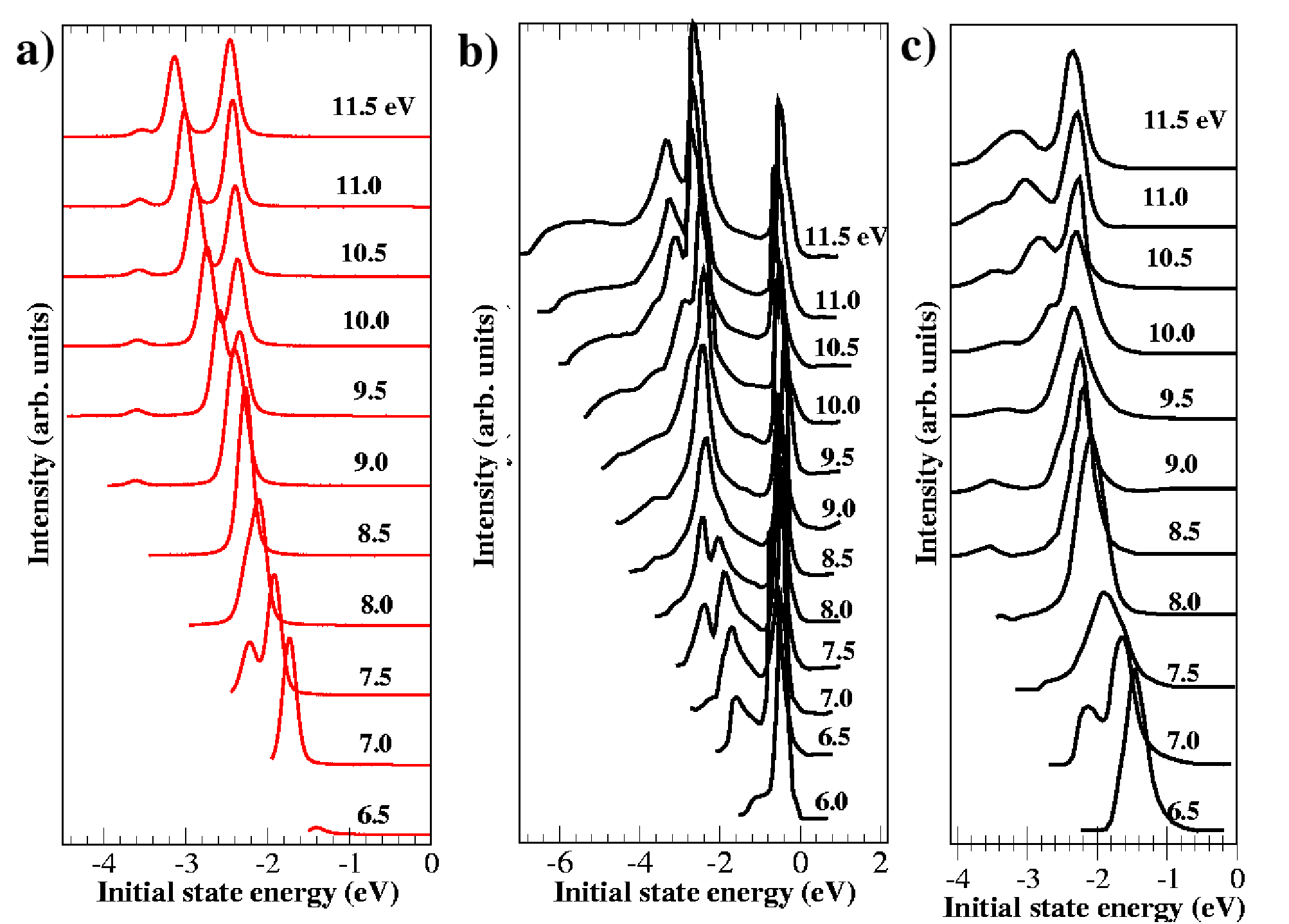}
    \caption{ (Color online)
      In panel a) we present our calculated photoemission
      spectra from Cu(111), while experimental
      spectra\cite{KnaHimEas79} are in panel b).
      Previous theoretical spectra \cite{SmiBenHur80} are in panel c).
    }
    \bigskip
    \label{fig:spectra111}
  \end{center}
\end{figure}

In Fig.~\ref{fig:spectra111} we compare our calculated,
experimental\cite{KnaHimEas79} and previous
theoretical\cite{SmiBenHur80} photoemission spectra from Cu(111) for
various photon energies indicated on the figure. Our calculated
spectra include the $\epsilon(\omega)$ and LDA+$U$-corrections.  The
angle of incidence was $60^\circ$ and the experiment was performed
with 90~\% $p$-polarized light. The previous calculations were done
within the three-step formalism,\cite{GroEasFre79} using an empirical
band structure generated by the combined-interpolation-scheme
approach.\cite{Blo62} In those calculations, the authors decided to additionally
suppress the $A_z$ contributions to the spectra, in order to get
better agreement with the experiment. Our intensities are not
adjusted, beyond the corrections imposed by Snell's law and Fresnel's
equations. Using the bulk-only model, we cannot reproduce the
surface peak present in the experimental spectra at about $-0.5$~eV, which is
missing also in the previous calculation.\cite{SmiBenHur80}
However, our spectra show overall similarity to the
experimental spectrum, especially considering the very general trends
of changing peak positions and intensities, with increasing photon
energy. In comparison with the previous calculation,\cite{SmiBenHur80} two 
spectra seem to be shifted by 0.5~eV in photon energy, {\em i.e.} 
our spectrum for $\hbar\omega=7.5$~eV is very similar to the one in 
Ref.~\onlinecite{SmiBenHur80} for 
$\hbar\omega=7.0$~eV.  Furthermore, the emergence of the first
bulk peak, originating from the $\lambda_1$ band 
at $\hbar\omega=6.5$~eV, (reduced in  intensity in our
calculation because of an overestimation of the work function), and its
shift to the deeper energies for increasing photon energy by 0.5~eV,
with a larger weight on the lower states, corresponds to the
experimental spectra. For $\hbar\omega=7$~eV, our spectrum has one $\lambda_1$
peak, while the transition from the $\lambda_3$ band at lower energy 
is not seen, because of the overestimated work function. Due
to poor experimental resolution, it is not easy to deduce if there are
two peaks or one in the experimental spectrum. The $\lambda_3$ peak is present
in our spectrum for $\hbar\omega=7.5$~eV, which reproduces well 
the experimental spectrum, regarding
both the peak positions and intensity. The previous calculation has
larger weight on the $\lambda_3$ peak, in contrast to the experiment. The
single peak in our spectra for $\hbar\omega=8-9$~eV shifts weight from
right to the left shoulder (higher to lower energy) and contains contributions
from both $\lambda_1$ and $\lambda_3$ initial states (which cross at those
energies). In experimental
spectrum, the same trend is present, but it starts from lower photon
energy (7.5~eV) and the spectra for lower photon energies have two
peaks.  For $\hbar\omega=9.0$~eV, a small peak at lower energy emerges ($\sim-3.1$~eV),
which corresponds to the $\lambda_3$ band and can be found also in all spectra for higher
photon energies. It is present also in the previous calculation and
in the experimental spectra, where it emerges at $\hbar\omega=8.5$~eV. 
Our spectra for $\hbar\omega=9.5-11.5$~eV have correct peak
positions, but wrong intensity ratio of peaks, significantly overestimating the
lower peak, originating from the $\lambda_1$-band, with
respect to the higher-energy peak, arising from the $\lambda_3$-band.
The previous calculations resolved this disagreement by artificially
suppressing the $z$-polarized contributions. A likely reason for this
disagreement is the neglect of surface effects in the three-step
model.

\section{Conclusions}

In this paper we have introduced a new method
for the {\em ab initio} low-energy photoemission
calculations based on the pseudopotentials and
plane-waves, which has an advantage in its simplicity and
unbiased basis-set, with the possibility to significantly
reduce the number of empirical parameters.
 Our method based on the bulk-emission model results in a
reasonable agreement with experiment in the photon energy
range up to $\sim25$~eV. Empirical corrections, including
the LDA+$U$ and surface-optics, give significant
improvements. Nevertheless, in comparison with the
one-step model, the intensity ratios of the
photoemission peaks are still not fully reproduced.
This is due to the neglect of surface damping
in our model, which, in principle, could be accounted
for within our approach. With respect to the experiment,
some broad structures are absent in our spectra,
which we interpret to originate from the forbidden
transitions in the normal photoemission, caused by the
detector's finite-acceptance angle and the related
broadening of ${\bf k_{\parallel}}$. Further work in this
direction, {\em i.e.} consideration of off-normal
photoemission, is necessary to assess these effects.

\acknowledgments
We acknowledge useful discussions with Ivana Vobornik, Alexander Smogunov, 
Nadia Binggeli and Paolo Umari. 

\bibliography{photoemission}

\begin{thebibliography}{59}
\bibitem[{PP_()}]{PP_energies}
\bibinfo{note}{The projector energies were: -0.39 and 4.20~eV for the
  $s$-channel, -0.05 and 2.00~eV for the $p$-channel and -0.49 and 2.80~eV for
  the $d$-channel.}
\bibitem[{not()}]{note_T}
\bibinfo{note}{We note that the $T(E_n, {\bf k})$ had to be interpolated, so
  that $E_i(k_\perp)$ and $T(E_n, {\bf k})$ are given on the same mesh of
  $k_\perp$ used for integration in Eq.~1.}
\end{thebibliography}
 
\end{document}